\documentclass[%
reprint,
superscriptaddress,
nofootinbib,
amsmath,amssymb,
aps,
pre,
]{revtex4-2}

\usepackage{graphicx}
\usepackage{epstopdf}
\usepackage{dcolumn}
\usepackage{bm}
\usepackage{hyperref}
\hypersetup{
     colorlinks=true,
     linkcolor=black,
     filecolor=black,
     citecolor = black,
     urlcolor=blue,
     }

\begin{document}

\title{Abnormal Critical Fluctuations Revealed by Magnetic Resonance in the Two-Dimensional Ferromagnetic Insulators}

\author{Zefang Li}
\affiliation{Beijing National Laboratory for Condensed Matter Physics, Institute of Physics, Chinese Academy of Sciences, Beijing 100190, China}
\affiliation{University of Chinese Academy of Sciences, Beijing 100049, China}

\author{Dong-Hong Xu}
\affiliation{Beijing National Laboratory for Condensed Matter Physics, Institute of Physics, Chinese Academy of Sciences, Beijing 100190, China}
\affiliation{University of Chinese Academy of Sciences, Beijing 100049, China}

\author{Xue Li}
\affiliation{Beijing National Laboratory for Condensed Matter Physics, Institute of Physics, Chinese Academy of Sciences, Beijing 100190, China}
\affiliation{University of Chinese Academy of Sciences, Beijing 100049, China}

\author{Hai-Jun Liao}
\affiliation{Beijing National Laboratory for Condensed Matter Physics, Institute of Physics, Chinese Academy of Sciences, Beijing 100190, China}
\affiliation{Songshan Lake Materials Laboratory, Dongguan, Guangdong 523808, China}

\author{Xuekui Xi}
\affiliation{Beijing National Laboratory for Condensed Matter Physics, Institute of Physics, Chinese Academy of Sciences, Beijing 100190, China}

\author{Yi-Cong Yu}\email{ycyu@wipm.ac.cn}
\affiliation{Beijing National Laboratory for Condensed Matter Physics, Institute of Physics, Chinese Academy of Sciences, Beijing 100190, China}
\affiliation{State Key Laboratory of Magnetic Resonance and Atomic and Molecular Physics, Wuhan Institute of Physics and Mathematics, Innovation Academy for Precision Measurement Science and Technology, Chinese Academy of Sciences, Wuhan 430071, China}

\author{Wenhong Wang}\email{wenhong.wang@iphy.ac.cn}
\affiliation{Beijing National Laboratory for Condensed Matter Physics, Institute of Physics, Chinese Academy of Sciences, Beijing 100190, China}%
\affiliation{Songshan Lake Materials Laboratory, Dongguan, Guangdong 523808, China}

\date{\today}

\begin{abstract}
    Phase transitions and critical phenomena, which are dominated by fluctuations and correlations, are one of the fields replete with physical paradigms and unexpected discoveries. Especially for two-dimensional magnetism, the limitation of the Ginzburg criterion leads to enhanced fluctuations breaking down the mean-field theory near a critical point. Here, by means of magnetic resonance, we investigate the behavior of critical fluctuations in the two-dimensional ferromagnetic insulators $\rm CrXTe_3 (X=Si, Ge)$. After deriving the classical and quantum models of magnetic resonance, we deem the dramatic anisotropic shift of the measured $g$ factor to originate from fluctuations with anisotropic interactions. The deduction of the $g$ factor behind the fluctuations is consistent with the spin-only state (${g\approx}$ 2.050(10) for $\rm CrSiTe_3$ and 2.039(10) for $\rm CrGeTe_3$). Furthermore, the abnormal enhancement of $g$ shift, supplemented by specific heat and magnetometry measurements, suggests that $\rm CrSiTe_3$ exhibits a more typical two-dimensional nature than $\rm CrGeTe_3$ and may be closer to the quantum critical point.
\end{abstract}

\maketitle

Fluctuations and correlations drive abundant phase transitions and critical phenomena. Regardless of the classical or quantum regime or of the order parameter and symmetry, they are universal in nature and follow statistical laws \cite{StaPhy}. Among them, a particularly fascinating aspect of two-dimensional (2D) magnetism associated with strong intrinsic magnetization fluctuations has introduced rich physical paradigms: the quantum spin liquid (QSL) state of the Kitaev model, Berezinskii-Kosterlitz-Thouless (BKT) transition of the XY model, Mermin-Wagner theorem of the isotropic Heisenberg model, Ising transition, etc. \cite{nature563}. Herein, recent discoveries of magnetic van der Waals (vdW) materials provide the ideal platform for exploring intrinsic 2D magnetism down to the 2D limit and potential opportunities for new spin-related applications \cite{Gongeaav4450}.

Notably, 2D magnetism is particularly susceptible to fluctuations. The Ginzburg criterion indicates that fluctuations become much more relevant with decreasing dimensions, leading to the failure of mean-field theory \cite{Zappoli2015}. The Mermin-Wagner theorem recognizes that no long-range order can survive thermal fluctuations at finite temperature in a 2D system with continuous symmetry \cite{PhysRevLett.17.1133}. However, by breaking the continuous symmetry, anisotropy in the exchange interaction will open up the spin wave excitation gap to resist thermal agitations of magnons at finite temperature. Such notable examples of magnetic order in single atomic layers have been discovered in $\rm CrI_3$ \cite{nature22391}, $\rm CrGeTe_3$ \cite{Nature7657}, $\rm Fe_3GeTe_2$ \cite{NMFei} and $\rm VSe_2$ \cite{nm10.1038}. Moreover, 2D magnetism is associated with strong intrinsic competition between quantum fluctuations and thermal fluctuations \cite{sachdev2007quantum}. In the ground state where thermal fluctuations vanish, the quantum fluctuations demanded by Heisenberg's uncertainty principle will dominate the quantum phase transition (QPT), which is driven by some nonthermal external parameters such as the magnetic field, pressure, or chemical doping \cite{YaTingJia97404}. At finite temperature, the energy of a system and the enthalpy of its thermal fluctuations compete, resulting in a classical phase transition (CPT). Although fluctuations play a crucial role in 2D magnetism, most of the theoretical predictions by ab initio methods are based on zero temperature and ignore fluctuations. Obtaining phase boundary information of fluctuations and critical points through experimental detection is very important.

Here, we demonstrate magnetization-fluctuation-induced effective $g$ factor anisotropy in the 2D ferromagnetic insulators ${\rm CrXTe_3 (X=Si, Ge)}$ by means of ferromagnetic resonance (FMR) and electron paramagnetic resonance (ESR). In general, the dominant critical fluctuations occur at the critical temperature $T_c$ and decay exponentially when deviating from $T_c$. Compared with $\rm CrGeTe_3$ (CGT), the observation of critical fluctuations with enhanced intensity and broad temperature range in $\rm CrSiTe_3$ (CST) is abnormal, which is associated with the 2D nature even in the bulk counterparts. Although the critical behavior can be indirectly characterized by neutron scattering \cite{Carteaux_1995,PhysRevB.92.144404}, magnetic susceptibility measurement \cite{PhysRevB.96.054406,RN176}, specific heat measurement \cite{Cheng_2005}, nuclear magnetic resonance \cite{Hu2020} and the dynamic magnetoelectric coupling technique \cite{chai2018observation} and directly characterized by real-time magneto-optical imaging technology \cite{NMJin}, accurately estimating the temperature dependence of magnetization fluctuations by means of magnetic resonance is very exciting.

\begin{figure}[ht]
    \includegraphics{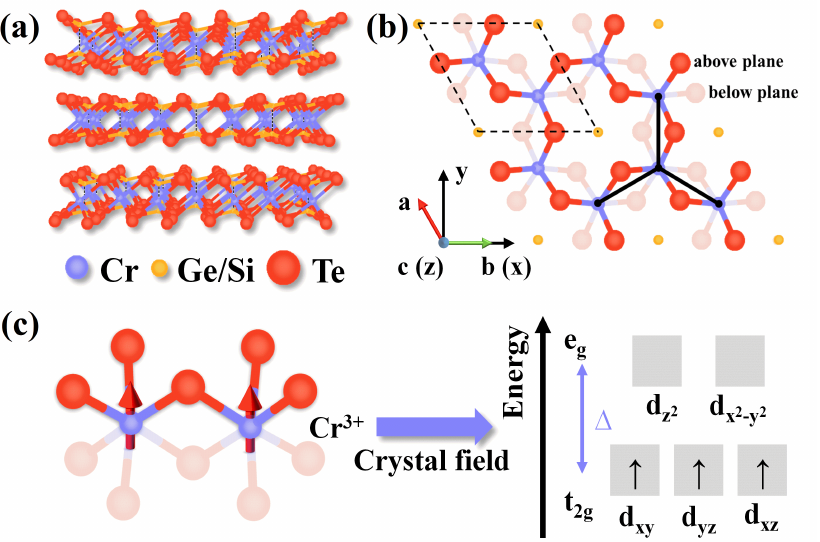}
\caption{\label{fig:str}(a) Crystalline structure of bulk rhombohedral $\rm CrXTe_3 (X=Si, Ge)$ with ABC vdW stacking. (b) Schematic of different coordinate systems from the top view of the ab plane: crystallographic axes a, b, and c and FMR coordinate axes x, y, and z. (c) Magnetic ion $\rm Cr^{3+}$ surrounded by a distorted octahedral crystal field with an out-of-plane arrow representing its easy magnetization direction. Split d levels in the $\rm e_g$ and $\rm t_{2g}$ manifolds, with 3 unpaired electrons in the $d_{xy}$, $d_{yz}$, and $d_{xz}$ orbitals.}
\end{figure}

2D vdW CST and CGT are ferromagnetic insulators and belong to the family of layered vdW transition metal trichalcogenides (TMTCs), which are crystallized in the $R\bar{3}$ (148) rhombohedral structure. Fig. \ref{fig:str} shows the honeycomb ABC layers stacked by a cleavable vdW gap. In each layer, the magnetic $\rm Cr^{3+}$ ions and $\rm Ge^{3+}/Si^{3+}$ pairs are located at $\rm Te^{2-}$ octahedral sites in a distorted $D_{3d}$ local symmetry, with a crystal field splitting the Cr-$3d^3$ orbitals and sustaining Cr-Te-Cr ferromagnetic superexchange. As shown in Fig. \ref{fig:str}(c), three unpaired electrons of $\rm Cr^{3+}$ are accommodated in the lower $t_{2g}$ triplet orbital, thus resulting in a quenched orbital moment for spins $J=S=3/2$ and a $g$ factor near 2.0023 for free electrons. Especially for CST, the strong Coulomb interactions from the narrow d bands and the half-filled condition favor a Mott transition \cite{PhysRevLett.123.047203}. As further shown in Fig. \ref{fig:str}(b), these octahedra arrange in edge-bond-sharing networks in the ab plane and form a magnetic honeycomb lattice. The interplay of spin-orbit coupling and the crystal field currently explains the uniaxial magnetic anisotropy with an easy axis perpendicular to the ab plane \cite{PhysRevLett.122.207201}. However, controversy about the uncertainty among the Kitaev, Ising, Heisenberg and single-ion anisotropy terms remains. CGT has been demonstrated to be well described by the Heisenberg behavior with a single-ion anisotropy term, which has been proven to exhibit ferromagnetic order in the monolayer \cite{Nature7657,PhysRevB.96.054406}. In contrast, CST with giant magnetic anisotropy is determined to be consistent with the 2D Ising behavior \cite{Carteaux_1995,RN176}, for which the ground state of the monolayer still lacks experimental confirmation. Moreover, in structurally related $\rm CrI_3$ \cite{PhysRevLett.124.017201,PhysRevB.101.134418}, $\rm \alpha-RuCl_3$ \cite{PhysRevLett.118.107203} and $\rm Na_2IrO_3$ \cite{PhysRevLett.110.076402}, Kitaev anisotropic exchange interactions are found in competition with Heisenberg interactions, which are associated with a possibly QSL state. Recently, first-principles-based simulations predicted the possible Kitaev QSL state in epitaxially strained CST monolayers \cite{PhysRevLett.124.087205,npj174}. After comprehensive consideration, we consider an XXZ Hamiltonian with single-ion anisotropy:
\begin{align}
    \label{eq:H}
    \begin{split}
    \mathcal{H} =&-\frac12\sum_{\langle j,l \rangle} (J \bm{S}_j \cdot \bm{S}_l+\Lambda S_j^zS_l^z)\\
    &-\sum_j A S_j^{z}S_j^{z}- \mu_{\rm B} \bm{H}\cdot \bm{g} \cdot \sum_j \bm{S}_j.
    \end{split}
\end{align}
The first term corresponds to the Heisenberg isotropic exchange $J$ and the anisotropic symmetric exchange $\Lambda$. The second term is the additional single-ion anisotropy term, and the last term corresponds to the Zeeman energy. $J>0$ favors ferromagnetic coupling, and $A>0$ favors the out-of-plane easy axis. Setting $\Lambda/J$ to infinity recovers the 2D Ising model, while the isotropic Heisenberg model is recovered for $\Lambda\approx0$ and $A\approx0$.

\begin{table}[ht]
    \caption{\label{tab:table1}%
    Physical quantities extracted from magnetometry and specific heat measurements \cite{sup}.}
    \begin{ruledtabular}
        \begin{tabular}{cccc}
                & \textrm{$\rm CrSiTe_3$} & \textrm{$\rm CrGeTe_3$}& \\
            \colrule
            Space group & $R\bar{3}(148)$ & $R\bar{3}(148)$ & \\ \hline
            Critical temp. & 34.15 & 68.15  & Derived MT \\
            $T_c$ $\rm (K)$ & 32.50& 65.50 &Arrott plot\\
                & 32.68 & 64.90 & Specific heat \\ \hline
            Curie-Weiss temp. & 57.16 & 101.46 & $H\parallel c$ \\
            $\Theta$ $\rm (K)$ & 53.86 & 100.24 & $H\parallel ab$ \\ \hline
            Frustration param. & 1.67 & 1.49 & $H\parallel c$\\
            $f$& 1.58 & 1.47 & $H\parallel ab$\\ \hline
            Effective mag.\footnote{Expected value $\mu_{{\rm eff}}=g\sqrt{(J(J+1))}{\rm \mu_B}=3.87\ {\rm \mu_B}$} & 4.00(5) & 4.03(4) & $H\parallel c$ \\
            $\mu_{{\rm eff}}$ $\rm (\mu_B)$ & 3.97(6) & 3.99(6) & $H\parallel ab$ \\ \hline
            Saturation mag.\footnote{Expected value $M_s=gJ{\rm \mu_B}=3.00\ {\rm \mu_B/f.u.}$} & 3.00 & 3.08 &$H\parallel c$ \\
            $M_s$ $\rm (\mu_B/f.u.)$ & 2.84 & 3.02 & $H\parallel ab$ \\ \hline
            Mag. entropy\footnote{Expected value $R\ln (2J+1)=11.53 \ {\rm J/mol \ K}$} & 3.91 & 0.86 & \\
            $\delta S$ $\rm (J/mol\ K)$ & 48.67\% & 38.85\% & above $T_c$ \\ \hline
            Critical exponent \footnote{$\rm CST$ close to 2D Ising model ($\beta=0.125, \gamma=1.75 $);\\ \noindent  $\rm CGT$ close to  tricritical mean-field model ($\beta=0.25, \gamma=1.0$)}& 0.171(6) & 0.221(2) & $\beta$ \\
                & 1.461(11) & 1.416(45) & $\gamma$ \\
                & 9.684(13) &7.287(12) & $\delta$ \\
        \end{tabular}
    \end{ruledtabular}
\end{table}

The presence of critical fluctuations near the critical point can be evidenced by magnetometry and specific heat measurements \cite{sup}. As shown in Table \ref{tab:table1}, Curie-Weiss behavior is observed at high temperatures, and an estimate of the effective moment gives $\mu_{{\rm eff}}\approx 4\ {\rm \mu_B}$, consistent with the spin-only magnetic moment $\mu_{{\rm eff}}=g\sqrt{(J(J+1))}{\rm \mu_B}=3.87\ {\rm \mu_B}$ for $J=S=3/2$ and $g=2.0023$. A deviation from the Curie-Weiss fit below $150\ \rm K$ can be recognized in the $M(T)$ curves, resulting in a higher Weiss temperature $\Theta$ than the critical temperature $T_c$. The ratio is defined as the frustration parameter $f=|\Theta|/T_c$, which corresponds to short-range ferromagnetic correlations persisting in the paramagnetic state. Moreover, the magnetic entropy $S(T)$ above $T_c$, estimated from the magnetic specific heat $C_m(T)$, recovers to a value as large as nearly $48.67\%$ (CST) and $38.85\%$ (CGT) of the total spin entropy. Such short-range correlations were indeed directly detected by elastic neutron scattering measurements in CST, as the measured correlation length remained larger than the nearest-neighbor distance up to 250 K \cite{PhysRevB.92.144404}. Furthermore, the critical exponents are obtained from modified Arrott plots extracted from the isotherm $M(H)$ curves, which well match the 2D Ising model for CST and the tricritical mean-field model for CGT. The above evidence implies that CST has a stronger critical fluctuation and an exchange interaction closer to a 2D nature, but further verification is needed.

\begin{figure}
    \includegraphics{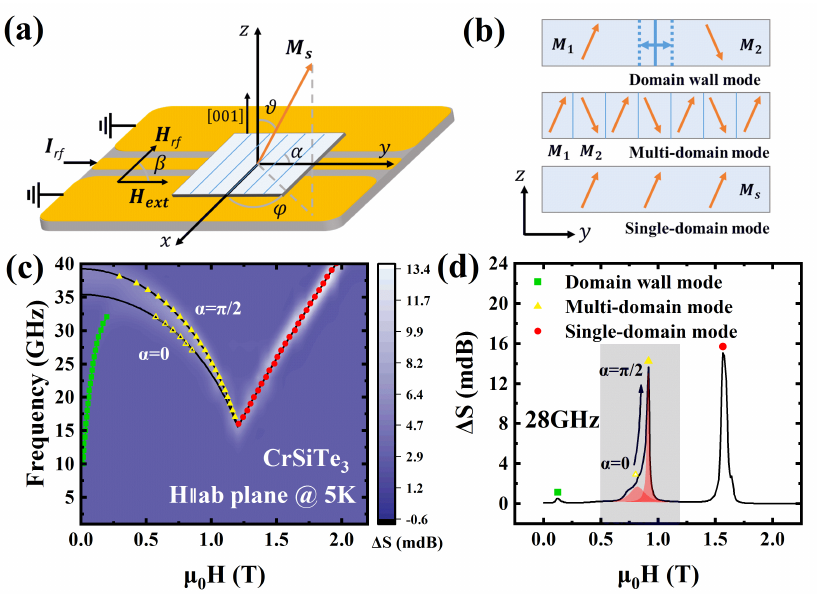}
\caption{\label{fig:fmr}(a) Coplanar waveguide with a rectangular single crystal placed parallel to the ab plane. The domain wall is represented by a blue solid line, and the angle with the y axis is $\alpha$. (b) Illustration of three different resonant modes from the side view. (c) Frequency- and field-dependent FMR spectra (in-plane, $\beta=\pi/2$) for $\rm CrSiTe_3$ at $\rm 5\ K$. The solid lines represent fitting for the single domain mode (fitted with Eq. \ref{eq:fmrr}) and multidomain mode (fitted with Eq. \ref{eq:fmrl}). (d) Typical microwave transmission at 28 GHz sliced from (c). The resonant peak of the multidomain mode is formed by the superposition of a series of peaks with different angles $\alpha$.}
\end{figure}

On the basis of the correspondence principle, the classical and quantum mechanical descriptions of the magnetic resonance are identical. Therefore, concrete expressions for the free energy and Hamiltonian are necessary to reveal the physics behind an observable. Here, we recover the resonance solutions of the Larmor equation (classical) and Heisenberg equation of motion (quantum) for a general case (more details are provided in the supplementary materials) \cite{sup,beljers1955ferromagnetic,JPSJ.43.857}. We conclude that the anomaly of the measured effective g factor, as well as the anomaly of the relationship between magnetocrystalline anisotropy $K_u$ and saturation magnetization $M_s$, can be explained by the specificity of fluctuations and correlations, which are unsettled in previously reported FMR measurements of $\rm CrI_3$ \cite{PhysRevLett.124.017201}, $\rm CrCl_3$ \cite{PhysRevLett.123.047204,PhysRevMaterials.4.064406}, and $\rm CrGeTe_3$ \cite{PhysRevB.99.165109,PhysRevB.100.134437,Zhang_2016}.

One of the important correspondences is the spectroscopic $g$ factor, which can be determined precisely by gyromagnetic ratio fitting using FMR and ESR spectra. As a classical description, shown in Fig. \ref{fig:fmr}(a), a rectangular-shaped single crystal is placed in a coplanar waveguide where it is acted upon by an alternating magnetic field ${H}_{\rm rf}$. In response to scanning of a strong homogeneous magnetic field ${H}_{\rm ext}$ at right angles, resonance absorption signals can be detected in the case that $\omega_{\rm res}=\gamma H_{\rm eff}$, where $\gamma=g\mu_{\rm B}/\hbar$ is the gyromagnetic ratio and $H_{\rm eff}$ is the effective internal field. When resonance occurs, the saturation magnetization $M_s$ induces Larmor precession along the effective field direction. In consideration of the Zeeman splitting energy, crystallographic anisotropy, demagnetizing field, and Bloch domain structure, the solutions to the Larmor equations are given by the Smit-Beljers approach \cite{sup,beljers1955ferromagnetic}:
\begin{widetext}
\begin{align}
    \label{eq:fmrr} & \left(\frac{\omega_{\rm res}}{\gamma}\right)^2=\left\{H-[H_A-(N_z-N_y)M_s]\right\}\left\{H-(N_y-N_x)M_s\right\},\\
    \notag & \text{for single-domain mode,} \ \vartheta_0=\varphi_0=\frac{\pi}{2},\ H \geq H_A+N_yM_s. \\
    \label{eq:fmrl} & \left(\frac{\omega_{\rm res}}{\gamma}\right)^2=(H_A+N_xM_s)(H_A+M_s\sin^2\alpha)-\frac{(H_A+M_s\sin^2\alpha-N_zM_s)(H_A+N_xM_s)}{(H_A+N_yM_s)^2}H^2, \\
    \notag & \text{for multidomain mode,}\ \varphi_{10}=\varphi_{20}=\frac{\pi}{2},\ \sin\vartheta_{10}=\sin\vartheta_{20}=\frac{H}{H_A+N_yM_s},\ H \leq H_A+N_yM_s.
\end{align}
\end{widetext}
where $H_A=2K/M_s$ is the anisotropic field, $N$ is the demagnetization factor, and $\alpha$ is the angle between the domain wall and the external magnetic field. When the magnetic field $H_{\rm ext}$ is applied parallel to the ab plane, we observe three different resonant modes (illustrated in Fig. \ref{fig:fmr}(b)), which are plotted as a function of excitation frequency and applied magnetic field in Fig. \ref{fig:fmr}(c). The domain wall resonance peaks excited under weak fields are much smaller in amplitude than the FMR peaks. Their dependence on the resonance frequency versus the in-plane field corresponds to the conventional theory for the Bloch wall model \cite{doi:10.1063/1.328926}. Remarkably, the multidomain mode has a continuously changing angle $\alpha$, resulting in asymmetric peak shapes (Fig. \ref{fig:fmr}(d)). The crossing point of the multidomain mode and single-domain mode indicates the saturation field of the domain structure ($H=H_A+N_yM_s$), which exists below the critical temperature $T_c$. Moreover, when the magnetic field $H_{\rm ext}$ is applied parallel to the c axis, the resonance frequency can be determined by the sum of the external field, the equivalent anisotropy field, and the demagnetizing field:
\begin{align}
    \label{eq:fmroop}
    \frac{\omega_{\rm res}}{\gamma}=H+H_A-N_zM_s, \ \text{for} \ \vartheta=0.
\end{align}

\begin{figure}[ht]
    \includegraphics{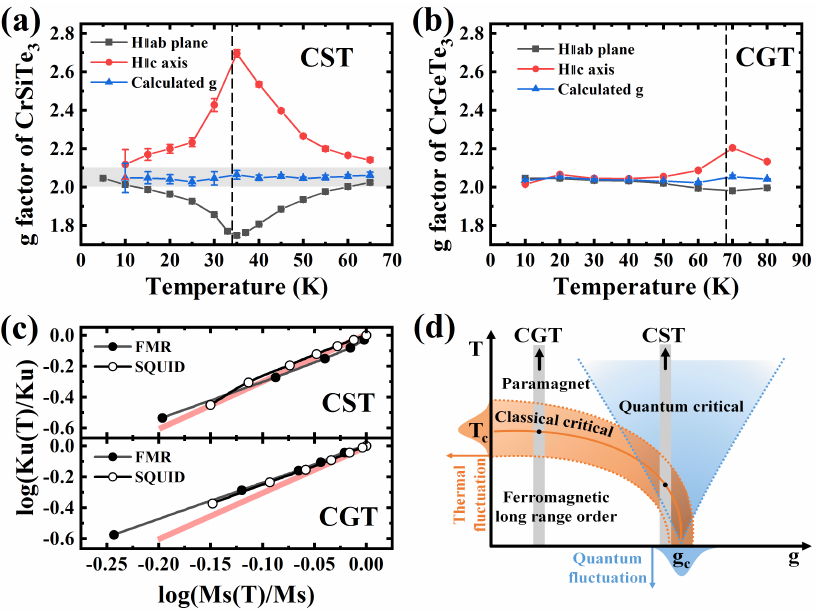}
\caption{\label{fig:cc}(a,b) Temperature dependence of the effective $g$ factor with the field applied both in-plane and out-of-plane for CST and CGT. The real $g$ factor is calculated with Eq. \ref{eq:g}. (c) Reduced anisotropy constant and magnetization at different temperatures shown on the logarithmic scale. The red line indicates an exponent of 3 for the Callen-Callen power law. (d) Schematic phase diagram showing the paramagnetic and ferromagnetic phases. The vertical paths for CST and CGT represent the CPT.}
\end{figure}

Herein, we extract the spectroscopic $g$ factor by fitting the in-plane and out-of-plane FMR and ESR data with the above equations (more details are provided in the supplementary materials) \cite{sup}. As shown in Fig. \ref{fig:cc}(a,b), the temperature dependence of the effective $g$ factor for the in-plane ($H \parallel ab$) and out-of-plane ($H \parallel c$) orientations has a contrasting behavior. A downwards (in-plane) or upwards (out-of-plane) shift of the $g$ factor is observed as the temperature increases, with a maximum value at $T_c$. Notably, the deviation of the $g$ factor is beyond an orbital contribution, which is almost completely quenched due to the crystal field \cite{PhysRev.178.497}. Such a temperature-dependent shift in the $g$ factor has been found in ESR measurements of low-dimensional metal alloys, metal complexes, or purely organic compounds \cite{PhysRevB.55.8398}. Based on Nagata's theory in the ESR case \cite{JPSJ.43.857,Nagata1977a,Nagata1972,Nagata1972a}, we strictly solve a general solution in the FMR case and conclude that magnetization fluctuations with anisotropic interactions are responsible for the $g$ shift (more details are provided in the supplementary materials) \cite{sup}. To be more specific, the Hamiltonian is substituted into the precession motion equation:
\begin{align}
    \label{eq:Nagata}
    \hbar \omega = \frac{\langle [S^{-},[S^{+},\mathcal{H}]] \rangle}{2\langle S^z \rangle}.
\end{align}
After calculating the thermodynamic average, we find that the isotropic Heisenberg term $J$ does not contribute to the $g$ shift, whereas the anisotropic symmetric exchange $\Lambda$ and single-ion anisotropy term $A$ do. The absolute value of the $g$ shift along the easy axis of magnetization $\Delta g_{c}$ is twice that along the orthogonal hard plane $\Delta g_{ab}$. By taking the value-weighted average, we can obtain the real $g$ factor of the sample:
\begin{align}
    \label{eq:g}
    g=\frac13 \times g_{ab}+\frac23 \times g_{c}.
\end{align}
As shown in Fig. \ref{fig:cc}(a,b), the calculated $g$ factor is a constant value independent of temperature. After averaging the $g$ factors over the entire temperature range, we obtain the averaged $g$ factors as 2.050(10) for CST and 2.039(10) for CGT, which are consistent with the orbital-quenched $g$ factors.

In addition, thermal fluctuations lead to an effective reduction in both the saturation magnetization and magnetocrystalline anisotropy, which can be represented by the Callen-Callen power law based on the single-ion anisotropy model:
\begin{align}
    \frac{K_u(T)}{K_u(0)}=\left[\frac{M_s(T)}{M_s(0)}\right]^{l(l+1)/2},
\end{align}
where $l$ is the order of spherical harmonics and depends on the symmetry of the crystal. In the case of uniaxial anisotropy for CST and CST, $l=2$ and an exponent of $3$ are expected. However, as shown in Fig. \ref{fig:cc}(c), CGT shows little agreement with the power law for exponents of 2.37(2) (FMR) and 2.51(3) (SQUID). In contrast, CST exhibits nonlinear behavior, which obviously violates the power law. Hence, the departure from the Callen-Callen power law suggests that the consideration of thermal fluctuations for single-ion anisotropy is incomplete. This is consistent with the fact that the single-ion anisotropy for $\rm Cr^{3+}$ is sufficiently small due to the weak spin-orbit coupling ($\xi \bm{L}\cdot\bm{S}$) with quenched orbital angular momentum ($L\approx 0$).

To illustrate the concepts, we consider the schematic phase diagram shown in Fig. \ref{fig:cc}(d), where $T$ is the temperature and $g$ is the strength of the ferromagnetic exchange coupling. On the one hand, the curve of the FM and PM phase boundary corresponds to the critical temperature $T_c$. The CPT occurs by varying the temperature through $T_c$. In the classical critical region, the correlation length tends to infinity, and critical fluctuations are dominant. On the other hand, changing $g$ in the ground state will lead to a QPT at the quantum critical point $g_c$, where the quantum fluctuations are the strongest. According to the results of our experiment, the critical temperature of CST is relatively low, and the fluctuations observed are much stronger than those for CGT. Therefore, we can reasonably indicate that CST is closer to the quantum critical point $g_c$, which is dominated by both classical and quantum critical behavior. This inference is also supported by a recent report on pressure-induced superconductivity in CST \cite{PhysRevB.102.144525}. We believe that doping, pressure, cleavage, and electrical regulation can achieve a QPT in CST, but more experimental verification is needed.

In summary, we have combined magnetic resonance, specific heat and magnetometry measurements to investigate the behavior of critical fluctuations in bulk CST and CGT single crystals. Although fluctuations near the critical temperature are natural in magnetic materials, the observation of such anisotropic shifts of resonance peaks in low-dimensional systems is unique because of the Ginzburg criterion. Despite the structural and electronic similarities, CST and CGT show strong contrasts in critical behavior. Our work implies the presence of short-range correlation far above $T_c$ and a signally 2D nature even in bulk counterparts of CST. Although CST shows a stronger magnetic anisotropy, the absence of ferromagnetic order in the monolayer should be attributed to the enhanced fluctuations. Last but not least, such unignorable magnetization fluctuations in 2D magnetic materials will interact with the spins of scatterers (X-rays, neutron beams, spin currents, etc.) and enhance the scattering effect. For the application of 2D magnetic materials in spintronic devices, the influence of magnetization fluctuations must be evaluated carefully.

This work was supported by the National Key R\&D Program of China (2017YFA0206303 and 2017YFA0303202), National Natural Science Foundation of China (11974406), and Strategic Priority Research Program (B) of the Chinese Academy of Sciences (CAS) (XDB33000000). We thank Prof. Shiliang Li and Dr. Wenshan Hong from the SC08 group at the Institute of Physics, Chinese Academy of Sciences, as well as Prof. Yisheng Chai from Chongqing University for useful discussion.

\bibliographystyle{apsrev4-2}
\providecommand{\noopsort}[1]{}\providecommand{\singleletter}[1]{#1}%
\end{document}


\title{{\mdseries Supplementary Materials for} \\ \quad \\ Abnormal Critical Fluctuations Revealed by Magnetic Resonance in Two Dimensional Ferromagnetic Insulators}

\author{Zefang Li}
\affiliation{Beijing National Laboratory for Condensed Matter Physics, Institute of Physics, Chinese Academy of Sciences, Beijing 100190, China}
\affiliation{University of Chinese Academy of Sciences, Beijing 100049, China}

\author{Dong-Hong Xu}
\affiliation{Beijing National Laboratory for Condensed Matter Physics, Institute of Physics, Chinese Academy of Sciences, Beijing 100190, China}
\affiliation{University of Chinese Academy of Sciences, Beijing 100049, China}

\author{Xue Li}
\affiliation{Beijing National Laboratory for Condensed Matter Physics, Institute of Physics, Chinese Academy of Sciences, Beijing 100190, China}
\affiliation{University of Chinese Academy of Sciences, Beijing 100049, China}

\author{Hai-Jun Liao}
\affiliation{Beijing National Laboratory for Condensed Matter Physics, Institute of Physics, Chinese Academy of Sciences, Beijing 100190, China}
\affiliation{Songshan Lake Materials Laboratory, Dongguan, Guangdong 523808, China}

\author{Xuekui Xi}
\affiliation{Beijing National Laboratory for Condensed Matter Physics, Institute of Physics, Chinese Academy of Sciences, Beijing 100190, China}

\author{Yi-Cong Yu}\email{ycyu@wipm.ac.cn}
\affiliation{Beijing National Laboratory for Condensed Matter Physics, Institute of Physics, Chinese Academy of Sciences, Beijing 100190, China}
\affiliation{State Key Laboratory of Magnetic Resonance and Atomic and Molecular Physics, Wuhan Institute of Physics and Mathematics, Innovation Academy for Precision Measurement Science and Technology, Chinese Academy of Sciences, Wuhan 430071, China}

\author{Wenhong Wang}\email{wenhong.wang@iphy.ac.cn}
\affiliation{Beijing National Laboratory for Condensed Matter Physics, Institute of Physics, Chinese Academy of Sciences, Beijing 100190, China}%
\affiliation{Songshan Lake Materials Laboratory, Dongguan, Guangdong 523808, China}

\date{\today}

\maketitle
\tableofcontents

\section{Crystal growth, Magnetometry and Specific heat}
\subsection{Methods}
High quality $\rm CrSiTe_3 \ (CST)$ and $\rm CrGeTe_3 \ (CGT)$ single crystals were grown by the self-flux method \cite{doi:10.1063/1.4914134,PhysRevB.96.054406}. The mixture of pure elements ($\rm Cr:Si:Te=1:2:6,\ Cr:Ge:Te=10:13.5:76.5$) were mounted in an alumina crucible, which was sealed inside a quartz tube under high vacuum ($\rm <10^{-4}\ Pa$). Then the quartz tube is heated up to $\rm 1100\ ^\circ C$ in a tube furnace and slowly cooled down to $\rm 700\ ^\circ C$. Excessive molten flux was centrifuged quickly before solidification. The final hexagonal flakes were shiny and soft, and were easy to peel off.

Energy dispersive X-ray spectroscopy (EDXS, equipped in Hitachi S-4800 microscope) was used for element analysis. As well as the X-ray diffraction (XRD, Bruker D8 Anvance) proved the stoichiometric ratio and high quality of CGT and CST single crystals. We also measured the heat capacity at zero field (PPMS-9T, Quantum Design physical properties measurement system) and characterized the temperature and field dependent magnetization (MPMS, Quantum Design magnetic property measurement system). In consideration of the demagnetization effect, it should be noted that the external applied field has been corrected for the internal magnetic field as $H_{int}=H_{ext}-NM$, where $N$ is the demagnetization factor \cite{doi:10.1063/1.367113} and $M$ is the measured magnetization.

\subsection{M(T) and M(H) curves}
\begin{figure*}
    \includegraphics{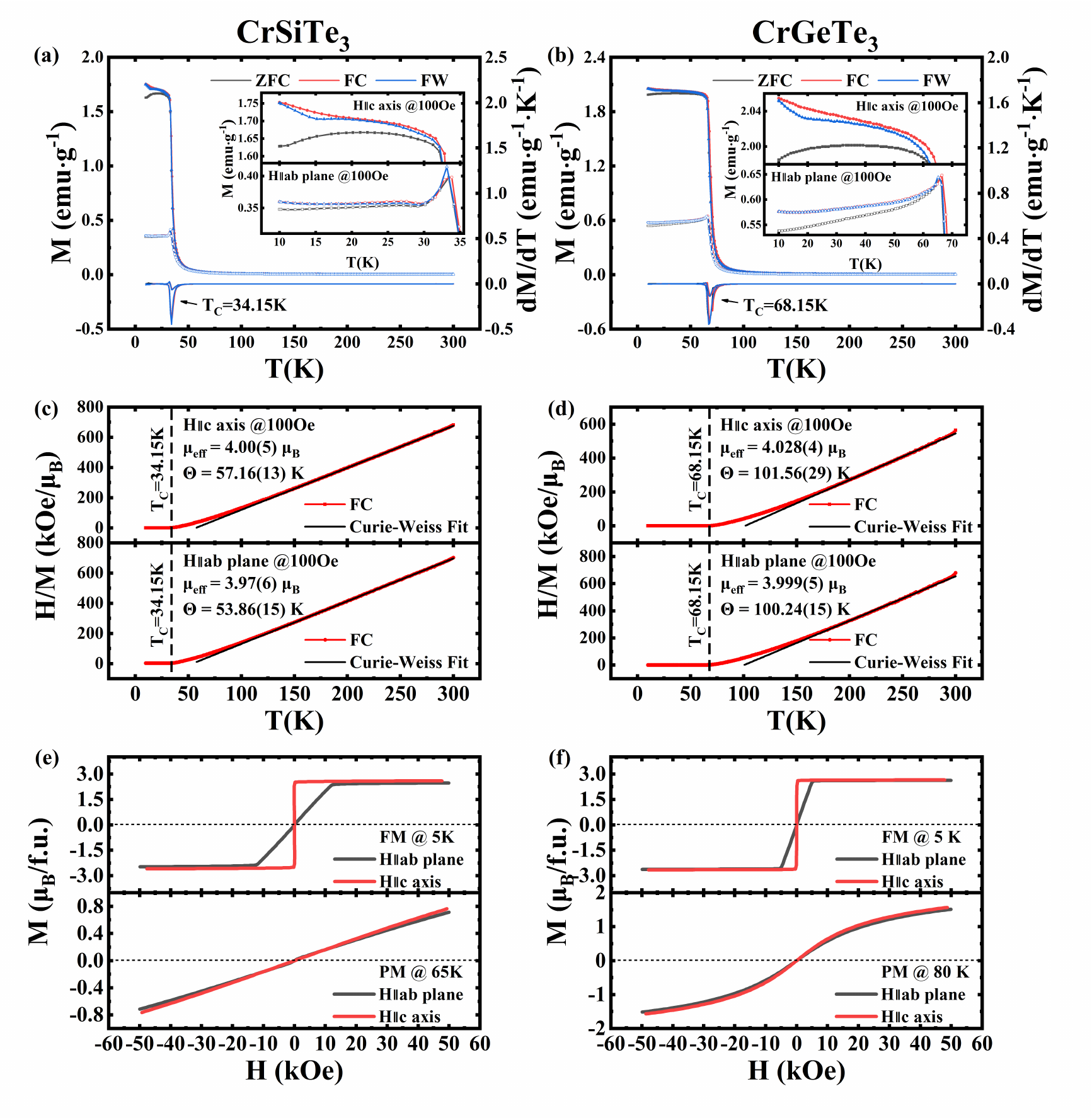}
    \caption{\label{fig:S1}(a), (b) Temperature dependence of the magnetization and its derivative measured in ZFC, FC and FW modes with a field of $\rm 100\ Oe$. The inset shows the enlarged picture of splitting between $H \parallel ab$ and $H\parallel c$. (c), (d) Inverse of the magnetic susceptibility for FC curve. The black solid lines indicate the linear fits in paramagnetic region with Curie-Weiss law. (e), (f) Magnetization as function of field at ferromagnetic and paramagnetic state respectively.}
\end{figure*}

The magnetometry results are listed in Fig. \ref{fig:S1}. Temperature dependent zero field cooled (ZFC), field cooled (FC), and field warmming (FW) with an external field of $H=100\ {\rm Oe}$ both parallel to c axis and ab plane are shown in Fig. \ref{fig:S1}(a) and Fig. \ref{fig:S1}(b). The paramagnetic-ferromagnetic (PM-FM) transitions occurs at the Curie temperature ($34.15\ {\rm K}$ for CST and $68.15\ {\rm K}$ for CGT) are determined by the derivative of M(T) curve, which are the highest among previously reported values \cite{doi:10.1063/1.4914134,RN176,PhysRevB.95.245212,PhysRevB.96.054406,PhysRevB.98.214420} and indicating a better sample quality. These two samples both exist a ``thermo-hysteresis" between FC and FW curves ($H\parallel c$) below $T_c$. As reported in $\rm CrGeTe_3$ \cite{doi:10.1021/acs.nanolett.9b02849} and $\rm Fe_3GeTe_2$ \cite{doi:10.1021/acs.nanolett.9b03453}, this anomalous behavior is related to the existence of skyrmions, but has not been confirmed in $\rm CrSiTe_3$. And then Fig. \ref{fig:S1}(c) and Fig. \ref{fig:S1}(d) display the temperature dependent inverse susceptibility H/M under FC, parallel to c axis and ab plane respectively. A linear fit of Curie-Weiss law at PM state yields the Weiss temperature $\Theta$ and effective moment $\mu_{\rm eff}$. Above $150\ K$, the fitted effective magnetic moment is close to the expected value $\mu_{{\rm eff}}=g\sqrt{(J(J+1))}{\rm \mu_B}=3.87\ {\rm \mu_B}$. Below $150\ K$, however, the H/M curves deviate from the linear fit, resulting a higher Weiss temperature $\Theta$ than critical temperature $T_c$. And the ratio of them is defined as frustration parameter $f=|\Theta|/T_c$. This suggests that short-range magnetic correlations persist in PM state above $T_c$. Fig. \ref{fig:S1}(e) and Fig. \ref{fig:S1}(f) show the field dependence of magnetization measured at FM and PM state. The M(H) curves splitting at 5K confirm the easy c-axis and larger magnetocrystalline anisotropy in CST than CGT. And the saturation magnetization fitted by the law of approaching saturation is very close to the expected value $M_s=gJ{\rm \mu_B}=3.00\ {\rm \mu_B/f.u.}$.

\subsection{Heat capacity}
\begin{figure*}[ht]
    \includegraphics{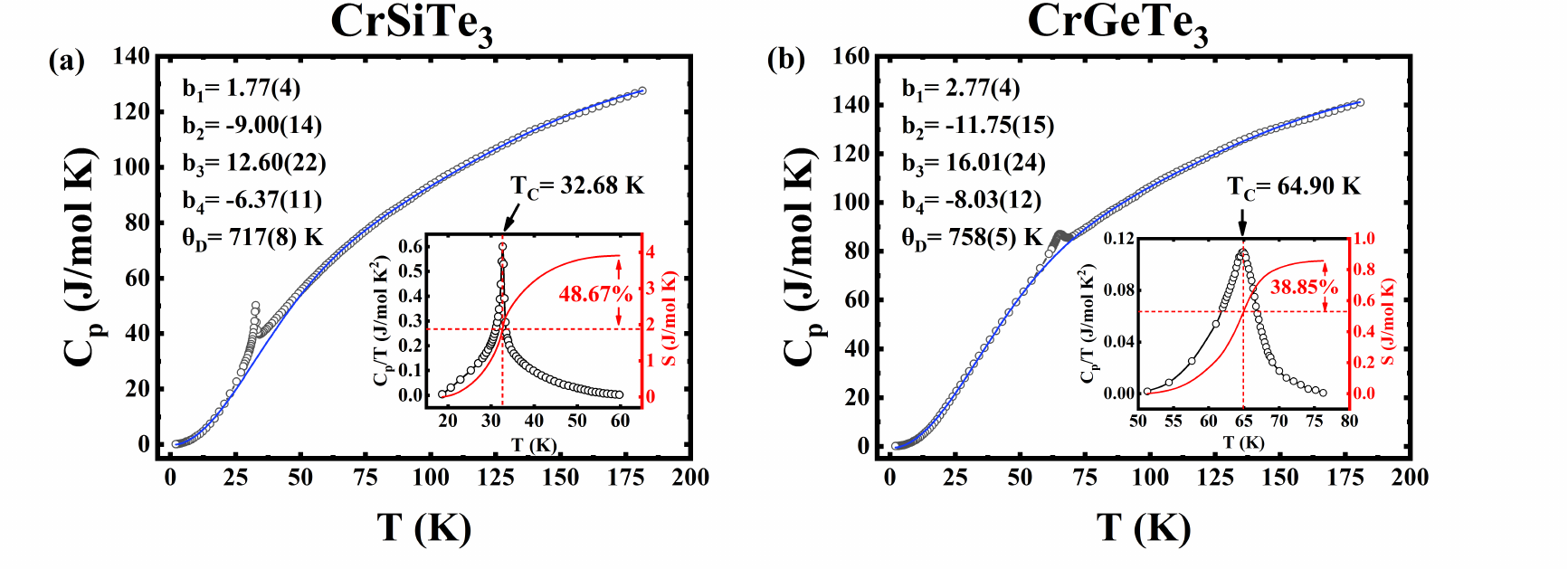}
    \caption{\label{fig:S2}(a), (b) Temperature dependence of zero-field specific heat for $\rm CrSiTe_3$ and $\rm CrGeTe_3$. The blue line is the fitting of lattice specific heat by Thirring model. The inset represents the magnetic contribution of Cp/T versus T and the integration of magnetic entropy S after subtracting the lattice contribution.}
\end{figure*}
Fig. \ref{fig:S2} shows specific heat data at zero field. As expected, the $\lambda$-shaped anomaly is observed at PM-FM transition. In order to compare the entropy change associated with magnetism, we subtract the non-magnetic contributions due to lattice vibrations. For this purpose, the lattice specific heat is estimated with the Thirring model \cite{Cheng_2005}:

\begin{align}
    \label{eq:Thirring}
    C_{\rm lattice}=3NR\left(1+\sum_{n=1}^\infty b_{n}\left(\left(\frac{2\pi T}{\theta_D}\right)^2+1\right)^{-n}\right).
\end{align}
where $N$ is the number of atoms in the unit cell, $R$ is the ideal gas constant, $\theta_D$ is the Debye temperature. We use $n$ up to $4$ for fitting and obtain a reasonable accuracy as shown by the blue solid line. After subtracting the lattice contribution, the inset shows the magnetic contribution $C_p/T$ versus $T$ and the magnetic entropy obtained from numerical integration. In general, the molecular field results in the second-order phase transition with a jump in specific heat. However, the fractions of the magnetic entropies gained above $T_c$ are $48.67 \%$ for CST and $38.85 \%$ for CGT. This behavior is attributed to short-range magnetic correlations between the moments of nearest-neighbor atoms.

At Zero field, the $2J+1$ energy states of $N$ non-interacting magnetic moments consist of $W=(2J+1)^N$ available states. Therefore the corresponding entropy can be calculated by Boltzman's theory:
\[
    S=k\ln W=Nk\ln(2J+1)=R\ln(2J+1)=11.53 {\rm J/mol \ K} \qquad \text{for} \qquad J=S=3/2.
\]
Due to short-range magnetic correlations, the measured entropy changes ($3.91\ \rm J/mol \ K$ for CST and $0.86\ \rm J/mol \ K$ for CGT) are relatively small.

\subsection{Arrott plot}
\begin{figure*}[ht]
    \includegraphics{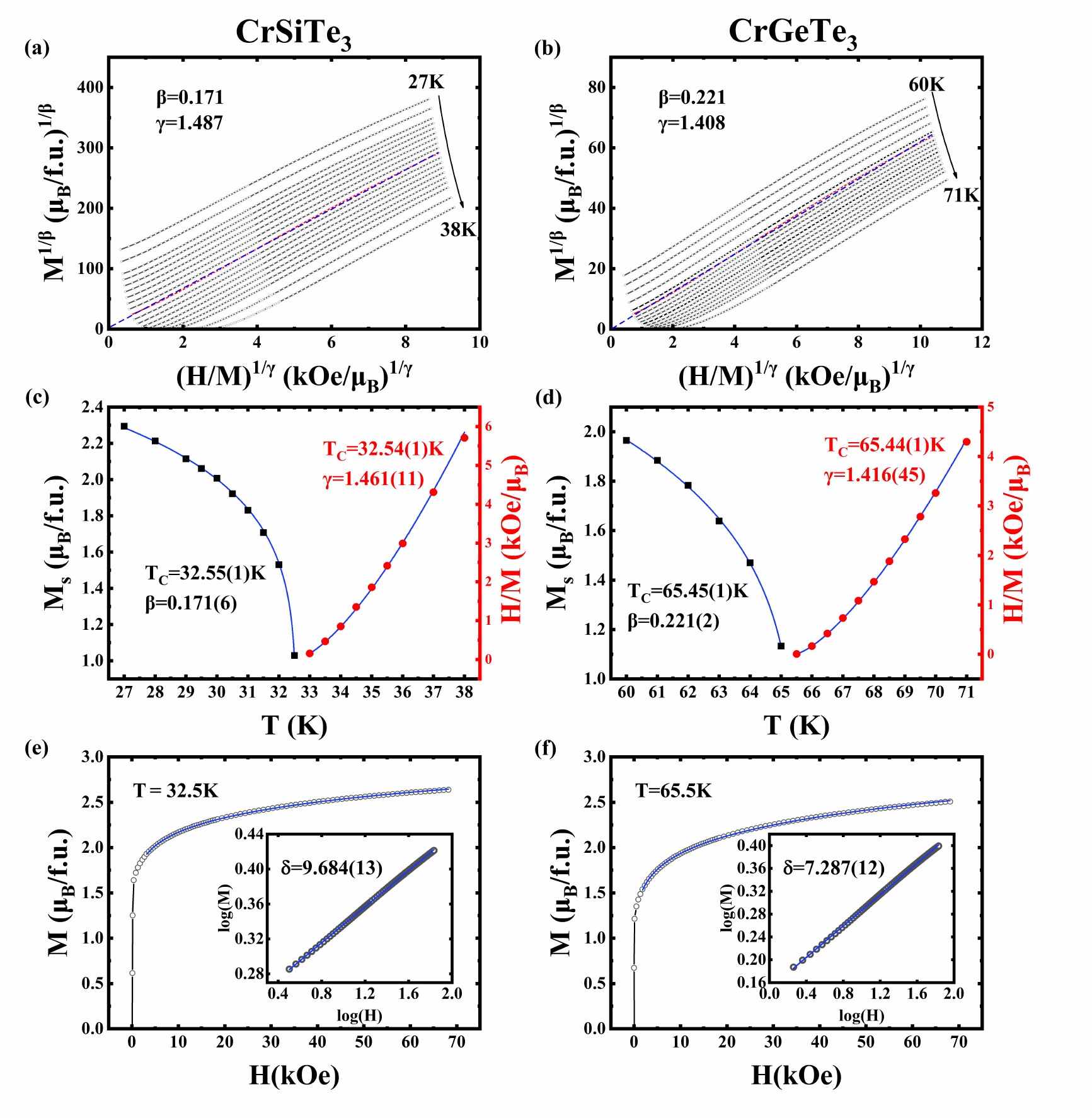}
    \caption{\label{fig:S3}: (a), (d) Modified Arrott plot of isotherms. The dashed line is the linear fit of isotherm at critical temperature. (b), (e) Temperature dependence of $M_s$ and $\chi$. The Tc and critical exponents are obtained from the fitting of Eq. \ref{eq:ex1} and \ref{eq:ex2}. (c), (f) Isothermal $MH$ at $T_c$. The inset shows the fitting of Eq. \ref{eq:ex3} and the extracted exponents.}
\end{figure*}

The second-order PM-FM phase transition near the critical point can be characterized by a set of critical exponents $\beta$, $\gamma$, and $\delta$. The spontaneous magnetization $M_s(T)$ below $T_c$, the inverse initial susceptibility $\chi_0^{-1}(T)$ above $T_c$ and the measured magnetization $M(H)$ at $T_c$ follow the power law:

\begin{align}
    \label{eq:ex1}
    M_s(T)\propto (T_c-T)^\beta, \quad T<T_c
\end{align}
\begin{align}
    \label{eq:ex2}
    \chi_0^{-1}(T)\propto (T-T_c)^\gamma, \quad T>T_c
\end{align}
\begin{align}
    \label{eq:ex3}
    M\propto H^{1/\delta},\quad T=T_c
\end{align}
It is known that the critical exponents are not independent of each other but follow the scaling relation:

\begin{align}
    \label{eq:scaling}
    \delta=1+\frac{\gamma}{\beta}
\end{align}
Therefore, we apply a modified Arrott plot by self-consistent iteration to determine the critical exponents and phase transition temperature $T_c$ \cite{PhysRevLett.19.786}. As shown in Fig. \ref{fig:S3}(a) and (b), isotherms plotted in form of $M^{1/\beta}$ versus $(H/M)^{1\gamma}$ constitute a set of parallel straight lines, and the isotherm at $T_c$ will pass through the origin. The relationship is given by the following equation:

\begin{align}
    \label{eq:arrott}
    \left(\frac{H}{M}\right)^{1/\gamma}=a\frac{T-T_c}{T}+bM^{1/\beta}
\end{align}
where $a$ and $b$ are constants. Linear fitting of the isotherms at high field region gives $(M_s)^{1/\beta}$ and $(\chi_0^{-1})^{1/\gamma}$ as an intercept on $M^{1/\beta}$ and $(H/M)^{1/\gamma}$. According to Eq. \ref{eq:ex1} and \ref{eq:ex2}, linear fitting of $\log[M_s(T)]$ versus $\log(T_c-T)$ and $\log[\chi_0^{-1}(T)]$ versus $\log(T-T_c)$ give new values of $\beta$ and $\gamma$, while free parameter $T_c$ is adjusted for best fitting. And then we substitute the new values into the Arrott plot and iterate repeatedly to get the optimal solution.

In order to test the accuracy of the final solution, we have analyzed the $(M_s)^{1/\beta}$ and $(\chi_0^{-1})^{1/\gamma}$ data by Kouvel-Fisher (KF) plot \cite{PhysRev.136.A1626}. As shown in Fig. \ref{fig:S3}(c) and (d), the values obtained are within the error accuracy. Furthermore, according to Eq. \ref{eq:ex3}, linear fitting of $\log[M(H)]$ versus $\log(H)$ gives a straight line with slope $1/\delta$. And it is noteworthy that these obtained $\beta$, $\gamma$, and $\delta$ are related in the scaling relation of Eq. \ref{eq:scaling}.

These obtained critical exponents for CST ($\beta=0.171(6), \gamma=1.461(11)$) can be explained by the 2D Ising model ($\beta=0.125, \gamma=1.75 $), while CGT ($\beta=0.221(2), \gamma=1.416(45)$) can be explained by tricritical mean-field model ($\beta=0.25$, $\gamma=1.0$). Our conclusions are consistent with with previous reports of neutron \cite{Carteaux_1995,PhysRevB.92.144404} and magnetic measurements \cite{RN176,PhysRevB.96.054406,PhysRevB.98.214420,PhysRevB.95.245212}.

\section{Derivation of Ferromagnetic Resonance Model and Experimental Fitting}
\subsection{Single-domain mode}

Considering a regular shaped magnetic sample in the uniform magnetic field, the motion of magnetization vector $\bf{M}$ follows the Larmor equation:
\begin{equation}
    \frac{d \bf{M}}{d t}=-\gamma
\left[
    \mu_0\bf{M}\times\bf{H}_{eff}
\right].
\end{equation}
The effective magnetizing field $\bf{H}_{eff}$ is determined by the free energy $F$ per unit volume. In single-domain case, it can be represented in the sum of Zeeman energy, the demagnetization energy, and magnetocrystalline anisotropy energy:
\begin{equation}
    F=-\mu_0M_sH\sin\vartheta\sin\varphi+K\sin^2\theta+\frac12\mu_0M_s^2(N_x\sin^2\vartheta\cos^2\varphi+N_y\sin^2\vartheta\sin^2\varphi+N_z\cos^2\theta).
\end{equation}
In which $K$ is the magneto-crystalline anisotropy, and $N$ is demagnetization factor. The equilibrium orientation of the magnetization vector $\bf{M}_s(\vartheta_0,\varphi_0)$ are determined by $\partial F/\partial \vartheta=0, \partial F/\partial \varphi=0$. When the vector $\bf{M}_s$ deviates slightly from the equilibrium position, the total free energy can be Taylor expanded as $F(\vartheta_0+\delta\vartheta,\varphi_0+\delta\varphi)=F(\vartheta_0,\varphi_0)+\frac12(F_{\vartheta\vartheta}\delta\vartheta^2+2F_{\vartheta\varphi}\delta\vartheta\delta\varphi+F_{\varphi\varphi}\delta\vartheta^2)$. In this case, the internal effective field are defined as $H_\vartheta=-(\mu_0M_s)^{-1}\partial F/\partial \vartheta, H_\varphi=-(\mu_0M_s\sin\vartheta_0)^{-1}\partial F/\partial \varphi$. Considering that Larmor precession has periodic solution $\delta\vartheta,\delta\varphi\propto e^{i\omega t}$, we obtain the equations of motion in the polar coordinate system as the following matrix form:
\begin{eqnarray}
    \left(
        \begin{array}{cc}
            F_{\varphi\vartheta}+i\omega\gamma^{-1}\mu_0 M_s\sin\vartheta_0 & F_{\varphi\varphi}\\
            F_{\vartheta\vartheta} & F_{\vartheta\varphi}-i\omega\gamma^{-1}\mu_0 M_s\sin\vartheta_0
        \end{array}
    \right)
    \left(
        \begin{array}{c}
            \delta\vartheta \\
            \delta\varphi
        \end{array}
    \right)=
    \left(
        \begin{array}{c}
            0 \\
            0
        \end{array}
    \right).
    \label{eq:e1}
\end{eqnarray}
Where $F_{\vartheta\varphi}=\partial^2 F/\partial \vartheta \partial \varphi$. The equation of motion has a non-zero solution only when the determinant is zero. Defining the anisotropic field $H_A=2K/M_s$, then we can get the resonance frequency as

\begin{align}
    \label{eq:fmrsl}
    \begin{split}
        \left(\frac{\omega_{\rm res}}{\gamma}\right)^2 =\left\{[H_A-(N_z-N_y)M_s]^2-H^2\right\}\frac{H_A-(N_y-N_x)M_s}{H_A-(N_z-N_y)M_s}, \\
        \text{for } \sin\vartheta_0=\frac{H}{H_A+(N_y-N_z)M_s}, \varphi_0=\frac{\pi}{2}, H< H_A+(N_y-N_z)M_s.
    \end{split}
\end{align}
\begin{align}
    \label{eq:fmrsh}
    \begin{split}
        \left(\frac{\omega_{\rm res}}{\gamma}\right)^2 =\left\{H-[H_A-(N_z-N_y)M_s]\right\}\left\{H-(N_y-N_x)M_s\right\}, \\
        \text{for } \vartheta_0=\varphi_0=\frac{\pi}{2}, H> H_A+(N_y-N_z)M_s.
    \end{split}
\end{align}
\subsection{Multi-domain mode}
For CST and CGT single crystals with hexagonal lattice and uniaxial anisotropy, the Bloch magnetic domain structure was observed experimentally below the saturation field \cite{doi:10.1021/acs.nanolett.9b02849,doi:10.1063/1.5024576}. It should be pointed out that Smit and Beljers first derived the multi-domain FMR model in $\rm BaFe_{12}O_{19}$ for a square-shaped sample. But a mistake in the derivation led to an unexplainable item in the result \cite{beljers1955ferromagnetic}. Here we recover the derivation and generalize to samples of arbitrary shape.

For simplicity, we assume that the magnetic domains of adjacent domains have two kinds of magnetization $\bf{M}_1(\vartheta_1,\varphi_1)$ and $\bf{M}_2(\vartheta_2,\varphi_2)$. The demagnetization energy of adjacent domains with infinitely thin thickness is equal to
\begin{align}
    \frac12\left(\frac{(\bf{M}_1-\bf{M}_2)\cdot \bf{e}_n}{2}\right)^2.
\end{align}
In considering the multi-domain case, the angle between the domain wall and the external magnetic field is $\alpha$, and the expression of free energy is equal to:
\begin{equation}
    \begin{split}
    F=&-\frac12\mu_0M_sH(\sin\vartheta_1\sin\varphi_1+\sin\vartheta_2\sin\varphi_2)+\frac12K(\sin^2\vartheta_1+\sin^2\vartheta_2)\\
    &+\frac12\mu_0M_s^2\left[\frac{N_x}{4}(\sin\vartheta_1\cos\varphi_1+\sin\vartheta_2\cos\varphi_2)^2+\frac{N_y}{4}(\sin\vartheta_1\sin\varphi_1+\sin\vartheta_2\sin\varphi_2)^2\right.\\
    &\left.+\frac{N_z}{4}(\cos\vartheta_1+\cos\vartheta_2)^2+\frac14(\sin\vartheta_1\cos(\varphi_1-\alpha)-\sin\vartheta_2\cos(\varphi_2-\alpha))^2\right].
    \end{split}
\end{equation}
Similarly, near the equilibrium position $\bf{M}_s(\vartheta_{10},\varphi_{10},\vartheta_{20},\varphi_{20})$, we can get the equation of motion in matrix form:
\begin{eqnarray}
    \left(
        \begin{array}{cccc}
            F_{11} & F_{12}-i\Omega/2 & F_{13} & F_{14} \\
            F_{21}+i\Omega/2 & F_{22} & F_{23} & F_{24} \\
            F_{31} & F_{32} & F_{33} & F_{34}-i\Omega/2 \\
            F_{41} & F_{42} & F_{43}+i\Omega/2 & F_{44}
        \end{array}
    \right)
    \left(
        \begin{array}{c}
            \delta\vartheta_1 \\
            \delta\varphi_1 \\
            \delta\vartheta_2 \\
            \delta\varphi_2 \\
        \end{array}
    \right)=\mu_0M_sH/2
    \left(
        \begin{array}{c}
            \cos\beta\cos\vartheta_{10} \\
            \sin\beta\sin\vartheta_{10} \\
            \cos\beta\cos\vartheta_{20} \\
            \sin\beta\sin\vartheta_{20}
        \end{array}
    \right).
\end{eqnarray}
Where $\Omega=\omega\gamma^{-1}\mu_0M_s\sin\vartheta_{10}$, $F_{12}=\partial^2 F/(\partial \vartheta_{10}\partial \varphi_{10})$, and $\beta$ is the angle between $\bf{H}_{rf}$ and $\bf{H}_{ext}$. In considering of the symmetry, we have $F_{11}=F_{33}$, $F_{22}=F_{44}$, $F{_{12}}=-F_{14}=-F_{23}=F_{34}$. For convenience we put $\Delta \vartheta^\pm=\delta\vartheta_{1}\pm\delta\vartheta_{2}$, $\Delta \varphi^\pm=\delta\varphi_{1}\pm\delta\varphi_{2}$. Then we get
\begin{eqnarray}
    \left(
        \begin{array}{cccc}
            F_{11}+F_{13} & -i\Omega/2 & 0 & 0 \\
            i\Omega/2 & F_{22}+F_{24} & 0 & 0 \\
            0 & 0 & F_{11}-F_{13} & 2F_{12}-i\Omega/2 \\
            0 & 0 & 2F_{12}+i\Omega/2 & F_{22}-F_{24}
        \end{array}
    \right)
    \left(
        \begin{array}{c}
            \Delta\vartheta^+ \\
            \Delta\varphi^+ \\
            \Delta\vartheta^- \\
            \Delta\varphi^- \\
        \end{array}
    \right)=\mu_0M_sH
    \left(
        \begin{array}{c}
            0 \\
            \sin\beta\sin\vartheta_{10} \\
            \cos\beta\cos\vartheta_{20} \\
            0
        \end{array}
    \right).
\end{eqnarray}
Therefore, the $4\times 4$ matrix can be split into two $2\times 2$ matrices, corresponding to two situations respectively:\\ (a) when $H_{rf}\perp H_{ext}\quad(\beta=\pi/2)$
\begin{equation}
    \label{eq:fmrmv}
    \begin{split}
        \left(\frac{\omega_{\rm res}}{\gamma}\right)^2=(H_A+N_xM_s)(H_A+M_s\sin^2\alpha)-\frac{(H_A+M_s\sin^2\alpha-N_zM_s)(H_A+N_xM_s)}{(H_A+N_yM_s)^2}H^2, \\
        \text{for}\quad \sin\vartheta_{10}=\sin\vartheta_{20}=\frac{H}{H_A+N_yM_s}, \varphi_{10}=\varphi_{20}=\frac{\pi}{2}.
    \end{split}
\end{equation}
(b) when $H_{rf}\parallel H_{ext}\quad(\beta=0)$
\begin{equation}
    \label{eq:fmrmp}
    \begin{split}
        \left(\frac{\omega_{\rm res}}{\gamma}\right)^2=(H_A+N_yM_s)(H_A+M_s\cos^2\alpha)-\frac{H_A+M_s\cos^2\alpha}{H_A+N_yM_s}H^2-M_s^2\sin^2\alpha\cos^2\alpha\left(1-\frac{H^2}{(H_A+N_yM_s)^2}\right), \\
        \text{for}\quad \sin\vartheta_{10}=\sin\vartheta_{20}=\frac{H}{H_A+N_yM_s}, \varphi_{10}=\varphi_{20}=\frac{\pi}{2}.
    \end{split}
\end{equation}

\subsection{Measured FMR spectra and experimental fitting}
\begin{figure*}[ht]
    \includegraphics{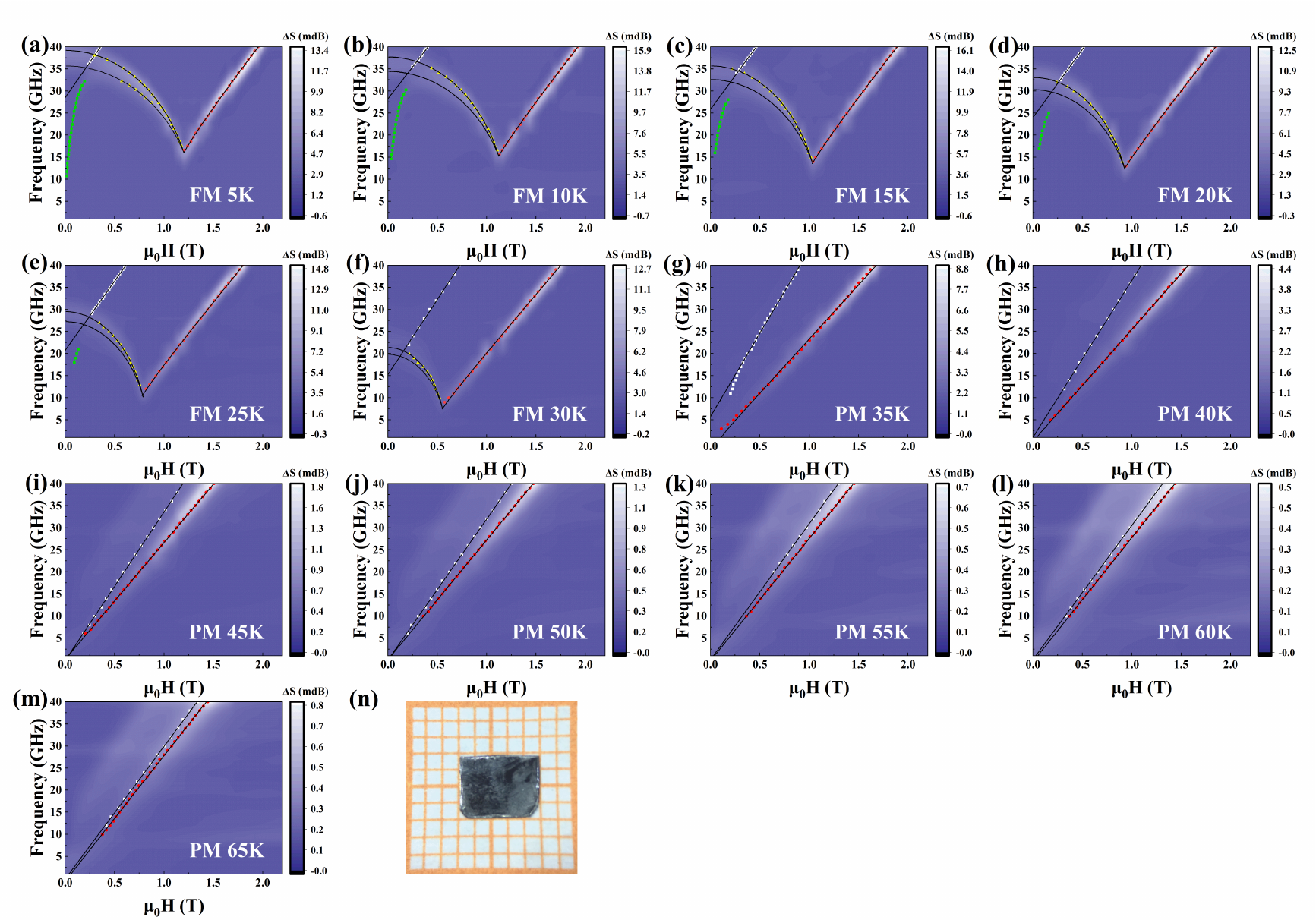}
    \caption{\label{fig:S4}: (a-m) Frequency and field dependent ferromagnetic resonance spectra for $\rm CrSiTe_3$. The color maps show the in-plane resonance spectra, while the white squares are the out-of-plane resonant peaks added afterwards. (n) Optical photograph of the single crystal used in the resonance experiment.}
\end{figure*}

\begin{figure*}[ht]
    \includegraphics{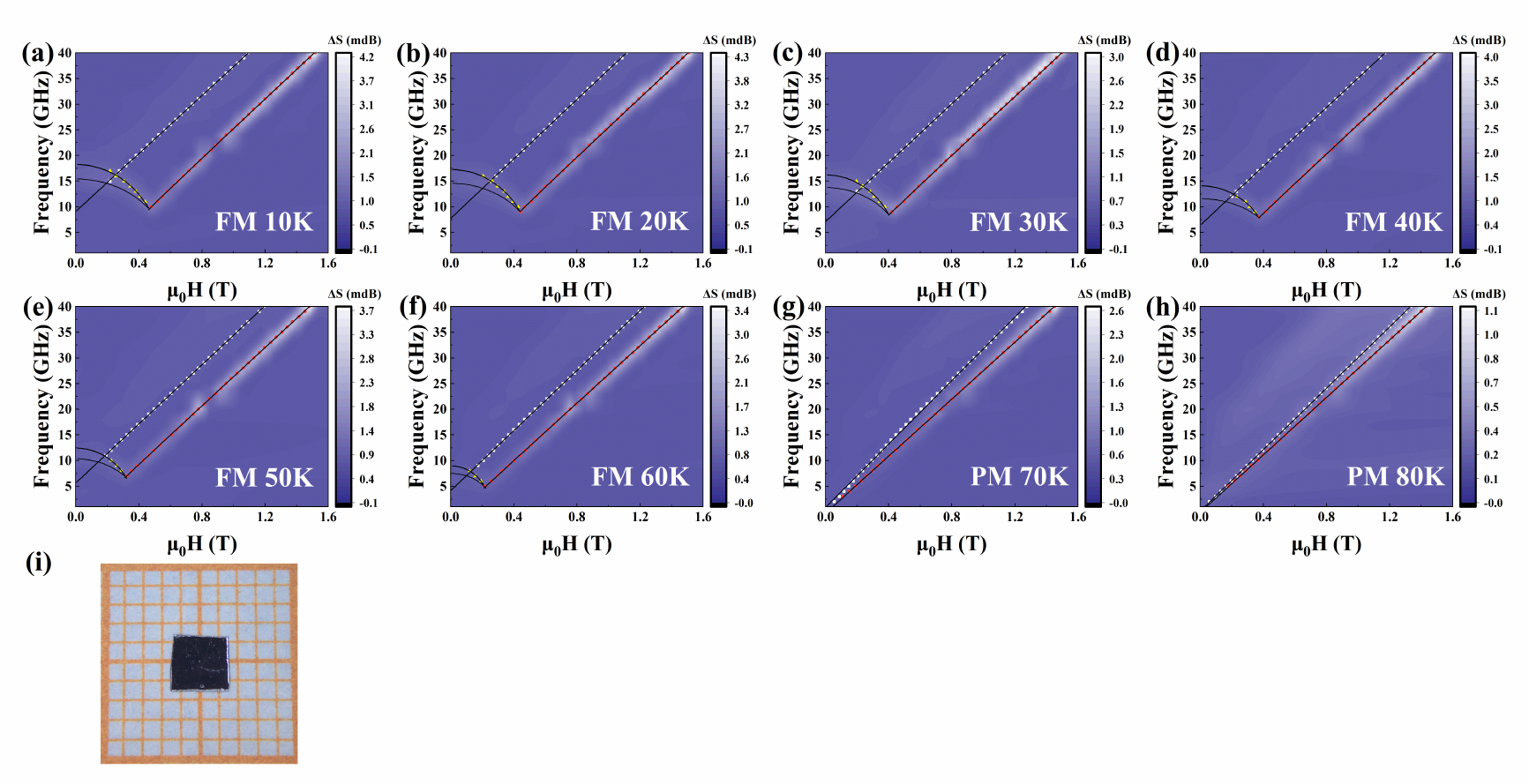}
    \caption{\label{fig:S5}: (a-h) Frequency and field dependent ferromagnetic resonance spectra for $\rm CrGeTe_3$. The color maps show the in-plane resonance spectra, while the white squares are the out-of-plane resonant peaks added afterwards. (i) Optical photograph of the single crystal used in the resonance experiment.}
\end{figure*}
As shown in Fig. \ref{fig:S4} and \ref{fig:S5}, broadband ferromagnetic resonance experiments were carried out on a home-made coplanar waveguide (CPW) sample rod, which was adapted to the magnetic field and temperature control system of Physical Property Measuring System (PPMS, Quantum Design). FMR spectra was recorded with a vector network analyzer (ZVA 40, Rohde \& Schwarz) in transmission mode (S12) over a frequency range of 1-40 GHz, in response to the scanning of magnetic field with the rate of 50 Oe per second. The resonance field was determined by Lorentzian fit to the spectra.

When the external magnetic field $\bf{H}_{ext}$ is applied in ab plane, we use Eq. \ref{eq:fmrsh} to fit the data above the saturation field ($H>H_A+N_yM_s$) and Eq. \ref{eq:fmrmp} to fit the data below the saturation field ($H<H_A+N_yM_s$). For $H=H_A+N_yM_s$, the two equations \ref{eq:fmrsh} and \ref{eq:fmrmp} get consistent results:

\begin{align}
    \label{eq:fmrcs}
    \left(\frac{\omega_{\rm res}}{\gamma}\right)^2=(H_A+N_xM_s)N_zM_s.
\end{align}
In order to determine these two free variables $\gamma$, $M_s$, we use equations \ref{eq:fmrmv} and \ref{eq:fmrcs} as two constraints to fit the data above the saturated field. Therefore, we can obtain the spectroscopic splitting factor $g$ from the formula $\gamma=g\mu_B/\hbar$, where $\mu_B=e\hbar/(2m_e)$ is the Bohr magneton.

When the external magnetic field $\bf{H}_{\rm ext}$ is applied in out-of-plane direction, the internal effective field $\bf{H}_{\rm eff}$ acting the spins is equal to the sum of the external field $H_{\rm ext}$, the equivalent anisotropy field $H_A=2K/M_s$, and the demagnetizing field $H_D=-N_zM_s$, whilst the resonant frequency can be determined from the well-known formula

\begin{align}
    \label{eq:fmroop}
    \frac{\omega_{\rm res}}{\gamma}=H+H_A-N_zM_s.
\end{align}

In the paramagnetic regime, usually the demagnetizing effects can be neglected as the $M_s$ is small. But for CST and CGT, obvious resonance peak shifts causes $\omega_{\rm res}-H$ curves to not exceed the zero point. Therefore, the g factor cannot be directly fitted by the formula of paramagnetic resonance $\omega_{\rm res}=\gamma H$. In order to eliminate the influence of the demagnetizing field, it is reasonable to fit the experimental data with equations \ref{eq:fmrmv} and \ref{eq:fmroop}. Because when $H_A=0$, $M_s=0$, and $N_x=N_y=N_z$ are satisfied with a total paramagnetic state, equations \ref{eq:fmrmv} and \ref{eq:fmroop} degenerate into $\omega_{\rm res}=\gamma H$. Therefore, our fitted g factor has excluded the influence of shape anisotropy.

\section{Quantum explanation of fluctuations induced g-factor anisotropy}
\subsection{Solution for a general model applicable for FMR and ESR case}
Consider a quasi-Heisenberg magnet consisting of similar spins with a common $g$-tensor. The Hamiltonian for such a spin system is expressed as \cite{Nagata1977}

\begin{align}
    \mathcal{H} = -2\sum_{\langle j,l \rangle} J_{jl} \bm{S}_j \cdot \bm{S}_l
         - \sum_{\langle j,l \rangle} \bm{S}_j \cdot \bm{\mathcal{K}}_{jl} \cdot \bm{S}_l
         - \mu_B \bm{H}\cdot \bm{g} \cdot \sum_j \bm{S}_j
         - A \sum_j (S_j^z)^2. \label{eqH}
\end{align}

Remark of the Hamiltonian (\ref{eqH}): (i) the first term is the usual Heisenberg exchanging coupling; (ii) the second term describes the magnetic dipole-dipole interaction, where the $\bm{\mathcal{K}}$ is a tensor and can contain non-diagonal elements; (iii) the third term is the Zeeman energy, $\bm{g}$ is the generalized $g$-factor which is also a tensor; (iv) the last term corresponds to the additional single-ion anisotropy \cite{Nagata1977a}.

The main technique to calculate the $g$-shift, which was developed by Nagata \cite{Nagata1972a}:

\begin{align}
    \hbar \omega = \frac{\langle [S^{-},[S^{+},\mathcal{H}]] \rangle} {2\langle S^z \rangle}.
    \label{eqFormula}
\end{align}
Here the $\omega$ is the frequency of the precession motion. Notice that in (\ref{eqFormula}) it is demanded that the direction of the effective magnetic field $\bm{H}$ is along $z$-axis (or saying, the axis of the precession motion is the $z$-axis). Substituting the Hamiltonian (\ref{eqH}) into the formula (\ref{eqFormula}), the frequency of precession can be evaluated one term by one term.

\paragraph{Zeeman term} First we evaluate the Zeeman term
$\mathcal{H}_{\text{Zeeman}} = - \mu_B \bm{H}\cdot \bm{g} \cdot \sum_j \bm{S}_j$.
Notice that this term has spatial translational symmetry, thus we only need to evaluate the single site, for this purpose we calculate

\begin{align}
    [S^{-},[S^{+},S^z]] = \frac{\hbar^3}{8}[X-\mathrm{i}Y,[X+\mathrm{i}Y,Z]]=2\hbar^2 S^z.
    \label{eqSz}
\end{align}
In the calculation above we take $S^x = \frac{\hbar}{2}X,S^y = \frac{\hbar}{2}Y,S^z = \frac{\hbar}{2}Z$ for convenience. Similarly

\begin{align}
[S^{-},[S^{+},S^x]]
&= \hbar^2(S^x - \mathrm{i}S^y) \notag \\
[S^{-},[S^{+},S^y]]
&= \mathrm{i}\hbar^2(S^x - \mathrm{i}S^y).
\label{eqSxSy}
\end{align}
The (\ref{eqSxSy}) terms are necessary when the off-diagonal $\bm{g}$ is take into consideration, although in all the Nagata's articles the $\bm{g}$ was always a diagonal tensor. For the off-diagonal case, we have

\begin{align}
    \bm{g} \triangleq
        \begin{bmatrix}
            g_{xx} & g_{xy} & g_{xz}\\
            g_{yx} & g_{yy} & g_{yz}\\
            g_{zx} & g_{zy} & g_{zz}
        \end{bmatrix}.
\end{align}
The external magnetic field vector is $\bm{H} = [H_x,H_y,H_z]$, then according to (\ref{eqSz}) and (\ref{eqSxSy}),

\begin{align}
    - \mu_B \bm{H}\cdot \bm{g} \cdot \bm{S}
    = &-\mu_B (g_{xx}H_x + g_{yx}H_y + g_{zx}H_z)S^x \notag \\
    & -\mu_B (g_{xy}H_x + g_{yy}H_y + g_{zy}H_z)S^y  \notag \\
    & -\mu_B (g_{xz}H_x + g_{yz}H_y + g_{zz}H_z)S^z,
\label{eqZeemanTerm}
\end{align}
then the contribution of the Zeeman term can be evaluated by (\ref{eqFormula})

\begin{align}
    &\hbar \omega_{\rm Zeeman}
    = \frac{\langle [S^{-},[S^{+},- \mu_B \bm{H}\cdot \bm{g} \cdot \bm{S}]] \rangle}
    {2\langle S^z \rangle} \notag \\
    & = -\mu_B \hbar^2(g_{xz}H_x + g_{yz}H_y + g_{zz}H_z)
     -\mu_B\frac{\hbar^2}{2}\left[(g_{xx}H_x + g_{yx}H_y + g_{zx}H_z)
     + \mathrm{i}(g_{xy}H_x + g_{yy}H_y + g_{zy}H_z)\right]
    \frac{\langle S^x - \mathrm{i}S^y \rangle}{\langle S^z \rangle}
\label{eqZeemanEnergy}
\end{align}
Notice that this term is different from the formula (6) in Nagata's article \cite{Nagata1977} with extra terms. Because in the Hamiltonian, the Zeeman term is negative defined, we have $-\mu_B H_z \hbar^2 g_{zz} = \mu_B \vert \bm{gH}\vert$, however, the $\langle S^x \rangle$, $\langle S^y \rangle$ and $\langle S^z \rangle$ in the extra term are hard to be determined. This difficulty comes from the off-diagonal element in the $\bm{g}$ tensor. The reason is, according to the original derivation in Nagata's article \cite{Nagata1972a}, the formula (\ref{eqFormula}) was arrived with assumption that \emph{``$S^{+}$ and $S^{-}$ are good normal modes"}, which seems to be not true if the $\bm{g}$ is not diagonal.

However, if the magnetic field was tuned above the saturation field in experiments, the inner effective magnetic field is parallel to the external field, then in this  case the formula (\ref{eqZeemanTerm}) and (\ref{eqZeemanEnergy}) can be simplified by $H_x = H_y = 0$, and moreover $\langle S^x \rangle = \langle S^y \rangle = 0$ because the magnetic precession is along the $z$-direction. In this respect we claim

\begin{align}
    \hbar \omega_{\text{Zeeman}} = -\mu_B \hbar^2 g_{zz} H_z.
    \label{eqZeemanEnergySimple}
\end{align}

\paragraph{Heisenberg term} $\mathcal{H}_{\text{Heisenberg}} = -2\sum_{\langle j,l \rangle} J_{jl} \bm{S}_j \cdot \bm{S}_l$ describes the isotropy exchanging interaction between near neighbour spins, and it is invariant under spatial rotations of the frame. The numerator of the (\ref{eqFormula}) is zero. To proof this, we evaluate

\begin{align}
    & [\mathcal{H}_{\text{Heisenberg}},\sum_i S_i^x + \mathrm{i} S_i^y ]
    \notag \\
    = & - \frac{\hbar^3}{4} \sum_{\alpha,i,\langle j,l \rangle} J_{jl}
    [\sigma_j^\alpha \sigma_l^\alpha, \sigma_i^x + \mathrm{i} \sigma_i^y]
    = -\frac{\hbar^3}{4} \sum_{\alpha,\langle j,l \rangle}
    J_{jl}[\sigma_j^\alpha \sigma_l^\alpha, \sigma_j^x + \mathrm{i} \sigma_j^y]
    + J_{jl}[\sigma_j^\alpha \sigma_l^\alpha, \sigma_l^x + \mathrm{i} \sigma_l^y]
    \notag \\
    = & - \frac{\hbar^3}{2} \sum_{\alpha,\beta,\langle j,l \rangle}
    \left( J_{jl}(\mathrm{i}\epsilon_{\alpha x \beta} \sigma_j^\beta \sigma_l^\alpha
    + \mathrm{i}\epsilon_{\alpha x \beta} \sigma_j^\alpha \sigma_l^\beta )
    - J_{jl}(\epsilon_{\alpha y \beta} \sigma_j^\beta \sigma_l^\alpha
    + \epsilon_{\alpha y \beta} \sigma_j^\alpha \sigma_l^\beta )
    \right)
    =0.
    \label{eqDeriveHeisenberg}
\end{align}
In (\ref{eqDeriveHeisenberg}) the $\epsilon$ is the antisymmetry tensor satisfying

\begin{align}
    \epsilon_{ijk} = \left\{
    \begin{array}{cc}
    1 & \quad \text{if } (ijk) = (xyz) \text{ or } (yzx) \text{ or } (zxy), \\
    -1 & \quad \text{if } (ijk) = (xzy) \text{ or } (zyx) \text{ or } (yzx), \\
    0 & \text{ otherwise. }
    \end{array} \right.
    \label{eqEpsilon}
\end{align}
The antisymmetry property (\ref{eqEpsilon}) leads to the last equal sign in the derivation (\ref{eqDeriveHeisenberg}). As a result, we have

\begin{align}
    \hbar \omega_{\text{Heisenberg}} = 0.
    \label{eqHeisenbergR}
\end{align}

\paragraph{Dipole-dipole interaction term} The dipole-dipole interaction

\begin{align}
\mathcal{H}_{\text{dipole}}
= - \sum_{\langle j,l \rangle} \bm{S}_j \cdot \bm{\mathcal{K}}_{jl} \cdot \bm{S}_l
= - \sum_{\langle j,l \rangle, \alpha, \beta} \mathcal{K}_{jl}^{\alpha \beta}
S_j^\alpha S_l^\beta
\end{align}
which was proposed in \cite{Nagata1977} is the generalization of the cases discussed in \cite{Nagata1972,Nagata1972a,Nagata1977a}. Here the $\bm{\mathcal{K}}$ is not necessary diagonal, although the diagonal case
(XXZ,XYZ model) is regarded to be quite general to describe the spin-spin interactions. The evaluation of the contribution of this term to the frequency is straight forward: substitute $\mathcal{H}_{\text{dipole}}$ into (\ref{eqFormula}), after tedious calculation we arrive at the result

\begin{align}
    \hbar \omega_{\rm dipole}
    = \sum_{\langle j,l \rangle, \alpha, \beta}
    T_{jl}^{\alpha \beta}
    \frac{\langle
    S_j^\alpha S_l^\beta \rangle}{2\langle S^z
    \rangle},
    \label{eqOmegaDipole}
\end{align}
and the form of the tensor $T_{jl}^{\alpha \beta}$ can be arranged into matrices

\begin{align}
    \bm{T}_{jl} = -\hbar^2
        \begin{bmatrix}
            -2\mathcal{K}_{jl}^{xx} + 2\mathcal{K}_{jl}^{zz}-2\mathrm{i}\mathcal{K}_{jl}^{xy}
            & -2\mathcal{K}_{jl}^{xy}+\mathrm{i}\mathcal{K}_{jl}^{xx}-\mathrm{i}\mathcal{K}_{jl}^{yy}
            & -5\mathcal{K}_{jl}^{xz}-\mathrm{i}\mathcal{K}_{jl}^{yz} \\
            -2\mathcal{K}_{jl}^{xy}+\mathrm{i}\mathcal{K}_{jl}^{xx}-\mathrm{i}\mathcal{K}_{jl}^{yy}
            & -2\mathcal{K}_{jl}^{yy} + 2\mathcal{K}_{jl}^{zz}+2\mathrm{i}\mathcal{K}_{jl}^{xy}
            & -5\mathcal{K}_{jl}^{yz}+\mathrm{i}\mathcal{K}_{jl}^{xz} \\
            -5\mathcal{K}_{jl}^{xz}-\mathrm{i}\mathcal{K}_{jl}^{yz}
            & -5\mathcal{K}_{jl}^{yz}+\mathrm{i}\mathcal{K}_{jl}^{xz}
            & 2\mathcal{K}_{jl}^{xx}+2\mathcal{K}_{jl}^{yy}-4\mathcal{K}_{jl}^{zz}
        \end{bmatrix}
\end{align}

Remark: When the external field is above the saturaction filed, the precession of the magnetics are along the $z$-axis and we could argue $\langle S^x S^y \rangle = \langle S^y S^z \rangle = \langle S^z S^x \rangle \approx 0$, and under this assumption we can safely drop the off-diagonal terms in (\ref{eqOmegaDipole}). Then we have

\begin{align}
    \hbar \omega_{\rm dipole}
    = \sum_{\langle j,l \rangle}
    \frac{\hbar^2}{\langle S^z \rangle}
    \left(
    (2\mathcal{K}_{jl}^{zz} - \mathcal{K}_{jl}^{xx} - \mathcal{K}_{jl}^{yy})\langle S_j^z S_l^z \rangle
    + (\mathcal{K}_{jl}^{xx} - \mathcal{K}_{jl}^{zz})\langle S_j^x S_l^x \rangle
    + (\mathcal{K}_{jl}^{yy} - \mathcal{K}_{jl}^{zz})\langle S_j^y S_l^y \rangle
    \right),
    \label{eqOmegaDipoleResult}
\end{align}
this corresponds with \cite{Nagata1977}. Notice that in the above derivation we ignore the term $K_{jl}^{xy}$ for the isotropy in the crystal plane. From (\ref{eqOmegaDipoleResult}) it can be read that

\begin{align}
    & \hbar \omega_{dipole}^\parallel
    = \sum_{\langle j,l \rangle}
    \frac{\hbar^2}{\langle S^z \rangle}
    (2\mathcal{K}_{jl}^{zz} - \mathcal{K}_{jl}^{xx} - \mathcal{K}_{jl}^{yy})
    \langle S_j^z S_l^z -S_j^x S_l^x\rangle
    \notag \\
    & \hbar \omega_{dipole}^\perp
    = \sum_{\langle j,l \rangle}
    \frac{\hbar^2}{\langle S^z \rangle}
    (\mathcal{K}_{jl}^{xx} - \mathcal{K}_{jl}^{zz})
    \langle S_j^z S_l^z -S_j^x S_l^x\rangle
    \label{eqOmegaDipoleResult2}
\end{align}

\paragraph{Single-ion anisotropy term} The single-ion anisotropy term $\mathcal{H}_\text{ani} =  A \sum_j (S_j^z)^2$ can be evaluated by

\begin{align}
    &[S^{-},[S^{+},\sum_j (S_j^z)^2]]
    = (\frac{\hbar}{2})^4 \sum_j[X_j - \mathrm{i}Y_j,[X_j+\mathrm{i}Y_j,Z_j^2]]
    = \sum_j \hbar^2 \left( 4(S_j^z)^2 - 2(S_j^x)^2 - 2(S_j^y)^2 \right)
    \notag \\
    &[S^{-},[S^{+},\sum_j (S_j^x)^2]]
    = (\frac{\hbar}{2})^4 \sum_j[X_j - \mathrm{i}Y_j,[X_j+\mathrm{i}Y_j,X_j^2]]
    = \sum_j \hbar^2 \left( 2(S_j^x)^2 - 2(S_j^z)^2  \right)
\end{align}
Then for the case that the external magnetic field is along $z$-axis, we arrive

\begin{align}
    \hbar \omega_\text{ani}^\parallel
    = \hbar^2 A \sum_j \frac{\langle 2(S_j^z)^2 - (S_j^x)^2 - (S_j^y)^2 \rangle}
    {\langle S^z \rangle}
    \label{eqOmegaAniParallel}
\end{align}
and when the external magnetic field is along $x$-axis

\begin{align}
    \hbar \omega_\text{ani}^\perp
    = \hbar^2 A \sum_j \frac{\langle (S_j^x)^2 - (S_j^z)^2\rangle}
    {\langle S^z \rangle}
    \label{eqOmegaAniPerp}
\end{align}

\paragraph{The thermodynamic average} Consider the finite-temperature case $\beta = \frac{1}{k_B T}$ in which the mean values of the operators can be evaluated by the partition function $Z = \mathrm{e}^{-\beta \mathcal{H}}$

\begin{align}
    \langle S^z \rangle
    =\frac{\mathrm{Tr}[\mathrm{e}^{-\beta H_0 - \beta H_f} S^z]}
    {\mathrm{Tr}[\mathrm{e}^{-\beta H_0 - \beta H_f}] }
    \simeq
    \frac{\mathrm{Tr}[\mathrm{e}^{-\beta H_0}(1-\beta H_f) S^z]}
    {\mathrm{Tr}[\mathrm{e}^{-\beta H_0}] }
\end{align}
where $H_f \doteq g\mu_BHS^z $ is the external magnetic filed, and $\mathcal{H} = H_0 + H_f$. Then we have

\begin{align}
    \langle S^z \rangle
    = -\beta g \mu_B H \langle\langle S^zS^z \rangle\rangle
\end{align}
where ``$\langle\langle \quad \rangle\rangle$" means average in zero filed. Similarly

\begin{align}
    &\langle S_j^z S_l^z \rangle
    = \beta^2 g^2 \mu_B^2 H^2 \langle\langle (S^z)^2 S_j^z S_l^z \rangle\rangle
    = \sum_{km} \beta^2 g^2 \mu_B^2 H^2
    \langle\langle S_k^z S_m^z S_j^z S_l^z \rangle\rangle
    \notag \\
    &\langle S_j^x S_l^x \rangle
    = \beta^2 g^2 \mu_B^2 H^2 \langle\langle (S^z)^2 S_j^x S_l^x \rangle\rangle
    = \sum_{km} \beta^2 g^2 \mu_B^2 H^2
    \langle\langle S_k^z S_m^z S_j^x S_l^x \rangle\rangle
\end{align}
Then according to formula (\ref{eqOmegaDipoleResult2}) we arrive at the expression

\begin{align}
    &\Delta \hbar \omega_\text{dipole}^\parallel
    = \frac{g_\parallel \mu_B H}{2k_B T \langle\langle S^z S^z\rangle\rangle}
    \sum_{jlkm} (\mathcal{K}_{jl}^{zz} - \mathcal{K}_{jl}^{xx})
    \langle\langle 2(S_j^z S_l^z - S_j^x S_l^x) S_k^z S_m^z\rangle\rangle
    \notag \\
    &\Delta \hbar \omega_\text{dipole}^\perp
    = \frac{g_\perp \mu_B H}{2k_B T \langle\langle S^z S^z\rangle\rangle}
    \sum_{jlkm} (\mathcal{K}_{jl}^{zz} - \mathcal{K}_{jl}^{xx})
    \langle\langle (-S_j^z S_l^z + S_j^x S_l^x) S_k^z S_m^z\rangle\rangle
    \label{eqTwiceDerivation}
\end{align}
and from the (\ref{eqTwiceDerivation}) we can obtain directly that

\begin{align}
    \frac{\Delta g_\text{dipole}^\parallel}{g_\text{dipole}^\parallel}
    = -2\frac{\Delta g_\text{dipole}^\perp}{g_\text{dipole}^\perp}.
    \label{eqTwiceDipole}
\end{align}
The similar argument can also arise for the results
(\ref{eqOmegaAniParallel}) and (\ref{eqOmegaAniPerp}), it can be proved

\begin{align}
    \frac{\Delta g_\text{ani}^\parallel}{g_\text{ani}^\parallel}
    = -2\frac{\Delta g_\text{ani}^\perp}{g_\text{ani}^\perp}.
    \label{eqTwiceAni}
\end{align}
Because the $g$-shift comes from the $\mathcal{H}_\text{dipole}$ and $\mathcal{H}_\text{ani}$, and the other terms in $\mathcal{H}$ do not contribute to the $\Delta g$, thus according to the formulas (\ref{eqTwiceDipole}) and (\ref{eqTwiceAni}), and we arrive

\begin{align}
    \frac{\Delta g^\parallel}{g^\parallel}
    = -2\frac{\Delta g^\perp}{g^\perp}.
\end{align}
We further assume that the $g$ factor is isotropic $g^\parallel=g^\perp=g$, and obtain $\Delta g^\parallel= -2\Delta g^\perp$, which is equivalent to

\begin{align}
    \label{eq:g}
    g=\frac13 \times g_{ab}+\frac23 \times g_{c},
\end{align}
where $g_{ab}=g+\Delta g^\perp$ and $g_{c}=g+\Delta g^\parallel$ are measured $g$ factor along the ab plane and c axis.

\subsection{Solution for the XXZ model with single ion anisotropy}
Consider a XXZ Hamiltonian with single ion anisotropy:

\begin{align}
    \label{eq:xxz}
    \mathcal{H} =-\frac12\sum_{\langle j,l \rangle} (J \bm{S}_j \cdot \bm{S}_l+\Lambda S_j^zS_l^z)
    -\sum_j A (S_j^{z})^2- \mu_{\rm B} \bm{H}\cdot \bm{g} \cdot \sum_j \bm{S}_j.
\end{align}
Based on the above derivation, we can directly draw conclusions: (i) the Heisenberg isotropic exchange $J$ has no contribution; (ii) the anisotropic symmetric exchange $\Lambda$, as well as the single-ion anisotropy $A$ contribute the $g$-shift.

\bibliographystyle{apsrev4-2}
\providecommand{\noopsort}[1]{}\providecommand{\singleletter}[1]{#1}%